# Iterative reconstruction of the detector response for medical gamma cameras

**A. Morozov,**[a,b,*] **V. Solovov,**[a] **F. Alves,**[c] **V. Domingos,**[a] **R. Martins,**[a,b] **F. Neves**[a,b] **and V. Chepel**[a,b]

[a] *LIP-Coimbra,*
*Department of Physics, University of Coimbra, Coimbra, Portugal*
[b] *Department of Physics,*
*University of Coimbra, Coimbra, Portugal*
[c] *ICNAS,*
*University of Coimbra, Coimbra, Portugal*
*E-mail*: andrei@coimbra.lip.pt

ABSTRACT: Statistical event reconstruction techniques can give better results for gamma cameras than the traditional centroid method. However, implementation of such techniques requires detailed knowledge of the PMT light response functions. Here we describe an iterative technique which allows to obtain the response functions from flood irradiation data without imposing strict requirements on the spatial uniformity of the event distribution. A successful application of the technique for medical gamma cameras is demonstrated using both simulated and experimental data. We show that this technique can be used for monitoring of the photomultiplier gain variations. An implementation of the iterative reconstruction technique capable of operating in real-time is also presented.

## 1. Introduction

Since its invention in 1957 by H.O. Anger [1,2], gamma camera remains an important and widely used molecular imaging tool in nuclear medicine [3-6], allowing to image the distribution of gamma ray-emitting radionuclides in a human body. In this type of detectors, a gamma ray interacts with a scintillation crystal and produces typically several thousands of optical photons which are detected by an array of photomultiplier tubes (PMTs). The position of a scintillation event can be found as the centroid of the PMT signals using analog electronics. Since in this method a non-linear response of the PMTs with the in-plane distance from the event position is approximated with a linear dependence, systematic distortions appear in the image, which are traditionally corrected by means of a look up table [3,4]. This approach, however, is very sensitive to the drift of the relative gains of the photomultipliers and requires regular re-calibrations of the camera to maintain high image quality.

An alternative approach to event position reconstruction is to use statistical reconstruction techniques (see e.g. [7-10]), based on finding the best match between the observed and expected PMT signals. These techniques have a number of advantages over the centroid reconstruction, such as potentially smaller distortions, larger useful field of view and better filtering of noise events [7,9]. However, statistical reconstruction requires detailed knowledge of the spatial response of the detector. In particular, the signal amplitude as a function of the scintillation event position should be known with high precision for each photomultiplier. We will further refer to these functions as Light Response Functions, or LRFs.

The difficulty in obtaining the spatial response of the detector with a sufficient precision constitutes the main obstacle for implementation of statistical position reconstruction methods for medical gamma cameras. Direct measurements (see e.g. [9,11]) or numerical simulations [12] can be used to establish the response. However, the direct measurements are very time consuming and typically need complicated robotics controls, while accurate simulations require knowledge of dozens of optical parameters.

Recently, a novel iterative approach for obtaining the detector response was developed for Anger-camera type detectors: a double phase liquid xenon detector for dark matter search [13] and a microstrip-based neutron detector with optical readout [14]. In this method, only flood field irradiation data are required to reconstruct the spatial response of the detector, which is performed using an iterative technique. There is no strict requirement on the irradiation uniformity, making acquisition of such data straightforward and quick.

Application of this technique for medical gamma cameras represents a number of challenges. This technique was originally developed for detectors in which light was produced in the so called secondary scintillation process: the atoms (or molecules) of the detector media were excited by impact of electrons accelerated in a strong electric field. The number of photons emitted in the secondary scintillation depends on the field strength and can be very large. In a medical gamma camera with a NaI(Tl) scintillation crystal, the number of photon produced by a 140 keV gamma ray is smaller by approximately two orders of magnitude, leading to potentially much larger statistical fluctuations in the PMT signals. Moreover, a larger contribution from the scattered and reflected light to the PMT signals can be expected for the crystal scintillators, which can strongly affect spatial symmetry of the gamma camera response, and, in turn, affect convergence of the method.

In this paper we demonstrate the feasibility of the iterative reconstruction technique for medical gamma cameras with both simulation and experimental data. A successful implementation of the technique is shown for a commercial gamma camera, retrofitted with a new readout system allowing to acquire signals of individual PMTs. We also introduce an open source package ANTS2, developed for simulation and experimental data processing for a broad range of Anger-camera type detectors and optimized for the iterative LRF reconstruction. Finally, we demonstrate real-time capabilities of our reconstruction technique by implementing calculations on a Graphics Processing Unit (GPU).

## 2. Iterative technique of LRF reconstruction

A scintillation event in a gamma camera is described by the position where the gamma photon is absorbed (XY coordinates) and the total number of generated optical photons which is approximately proportional to the deposited energy. A medical gamma camera is typically equipped with up to 100 photomultipliers. Several tens of amplitude values are thus recorded for each scintillation event in order to evaluate only three parameters. The redundancy of the acquired data can, in principle, be used to extract some information about the detector itself, namely to characterize the spatial response. Given a large set of events distributed over the entire field of view of the detector (not necessarily uniformly), it is possible to reconstruct the LRFs of the photomultipliers, at least in some cases as demonstrated in [13,14].

In these studies, the reconstruction of the detector response was performed with a method referred to as *adaptive reconstruction* using flood field illumination data. In short, the method consists in the following iterative procedure. Starting with an initial guess on the light response functions, these LRFs are used to reconstruct the spatial distribution of the scintillation events with one of the statistical reconstruction methods. In turn, the reconstructed event positions and the measured PMT signals are used to evaluate the LRFs. The iterations are repeated until the convergence of the LRFs is achieved.

The initial estimate for the LRFs can be obtained using either experimental or simulation data. In the case of simulations, the exact positions of the events are known and the PMT signals can be used directly to evaluate the LRFs. For experimental data, the centroid reconstruction can be used to calculate the approximate positions of the events. Note that the iterative LRF reconstruction has a substantial tolerance to the errors in the initial guess as shown in [14].

During the iterations, regularization of the LRFs has to be performed. For axially-symmetric LRFs, a pooling procedure was previously applied [13,14] to ensure that the obtained LRFs are monotonically decreasing functions. This property of the light response functions is expected from the radial dependence of the solid angle, subtended by the PMT photocathode at the light emission position. Without this regularization, the statistical reconstruction often produces artifacts (areas with very high or, the opposite, very low event density) which can lead to failure of the iterative procedure.

Another important factor is efficient discrimination of "bad" events such as the events with distorted signal values, double events or events from the detector periphery with a large error in the reconstructed position. This can be achieved by introduction of event filters, which, for example, limit the range of acceptable reconstructed energy or establish a cut on the reconstructed spatial coordinates.

Several techniques can be used during the iterative LRF reconstruction to improve the convergence speed or, in some cases, to avoid local minima. The first technique consists in organizing the PMTs in one or several groups, and using a common LRF for all the PMTs in a group to exploit spatial symmetry of the detector response. In this case a scaling factor for each individual PMT in a group has to be introduced to account for differences in the relative gains. For a detector with only one type of PMTs a good strategy is, typically, to perform several first iterations with a common LRF for all the PMTs, and, after the general shape is established, to use individual LRFs in the following iterations.

Another technique involves so called *passive* PMTs: the signals of the photomultipliers, which were declared *passive* by the user, are not used during the next iteration in position reconstruction. In this way, disabling PMTs with less-adequate LRFs one uses their neighbors to perform a more accurate reconstruction in the regions where reconstruction would otherwise be distorted, and, in turn, to improve the LRFs of these passive PMTs when the new LRFs are calculated.

Finally, a "blurring" procedure can also be useful: after the reconstruction was performed for all events, every reconstructed position is shifted randomly by a small value. Since LRFs are typically slow-changing functions, this procedure does not introduce large errors while it allows to populate the regions with no events due to inadequate shape of the LRFs in the current iteration.

## 3. Validation with simulation data

This section starts with a brief description of the simulation package used in this study. Then we show a successful application of the iterative LRF reconstruction technique for the data obtained in simulations of a medical gamma camera. Finally, we discuss how the flood field illumination data can be used to evaluate the effective center positions of the PMTs to take into account, for example, the photocathode non-uniformity.

### 3.1 ANTS2 package

A custom package ANTS2 has been developed to perform detector simulations, reconstruct scintillation events, obtain detector response from flood field irradiation data and process experimental data for Anger-camera type detectors. The package source code and a user guide are public and can be found in [15]. A detailed description of the package will be presented elsewhere, while here we provide only a general overview.

The package has three main units: simulation, event reconstruction and LRF calculation modules. The simulation module is designed to perform Monte Carlo simulations of scintillation events in Anger-camera type detectors with fully-customizable 3D detector geometry. CERN ROOT [16] libraries are used to store the detector geometry, provide 3D navigation during photon and particle tracking and to handle 3D visualization. Photons are traced according to the optical properties of the materials defined by the user. User-specified overrides can be introduced for material borders to take into account photon scattering. A detailed model of light detection by PMTs and pixelated semiconductor sensors is implemented in the package.

The event reconstruction module calculates the spatial coordinates and energy of individual events using sensor signals, assuming that the light emission originates from a point source. This module can perform event reconstruction for simulated or imported experimental data using one of several available statistical reconstruction algorithms based on evaluation of the least squares or likelihood of the reconstruction. One of the algorithms is implemented on GPU enabling real-time event reconstruction (see section 5). The module also features a large set of event filters and tools for analysis of the reconstruction quality.

The LRF module calculates and stores LRFs of the optical sensors. The LRFs are parameterized using cubic B-splines [17] according to several schemes exploiting the spatial symmetry of the detector response. The module allows to define several sensor groups and use a common LRF, combined with individual scaling factors (relative gains) within a group. Given a set of events, the module performs reconstruction of the LRFs (and relative gains, if necessary) using the chosen parameterization and grouping schemes. For simulated events, the module can use true or reconstructed positions. This option allows to compare the LRFs, reconstructed using the iterative procedure, with those determined from the simulation using the known true positions of the events. Below, we refer to the latter type as *actual* LRFs.

The package is optimized to perform the iterative LRF reconstruction using the technique described in section 2. A scripting tool is provided which allows to configure and run the iterations. Note that the implemented LRF parameterization offers sufficient LRF regularization, and no additional regularization techniques are required.

### 3.2 Simulation of a medical gamma camera

Simulations of a medical gamma camera were performed with the configuration shown in figure 1. This configuration represents a model of the commercial gamma camera used in the experiments described in section 4.2. In this configuration an array of 37 hexagonal PMTs is coupled to a Ø470 mm NaI(Tl) scintillation crystal through a Ø500 mm light guide. The crystal and light guide have the same thickness of 12.5 mm.

The crystal was irradiated uniformly with a parallel beam of gamma rays from a $^{57}$Co source (122 and 136 keV) through a mask comprised of lead bars. The bars were 10 mm wide and 3 mm thick, placed with a 20 mm step as shown in figure 1. The irradiated area of the crystal had a diameter of 400 mm corresponding to the dimensions of the standard collimator used in the experiments. Since no information has been found on the optical properties of the interface between the crystal and the encapsulation container (the back surface and side walls of the crystal), an assumption was made that 75% of the photons arriving at the interface are scattered back in $2\pi$ according to the Lambert's cosine law with the rest being absorbed. The relative gains of the PMTs were chosen randomly (uniform distribution from 0.25 to 1) to simulate uncertainty in the PMT gains.

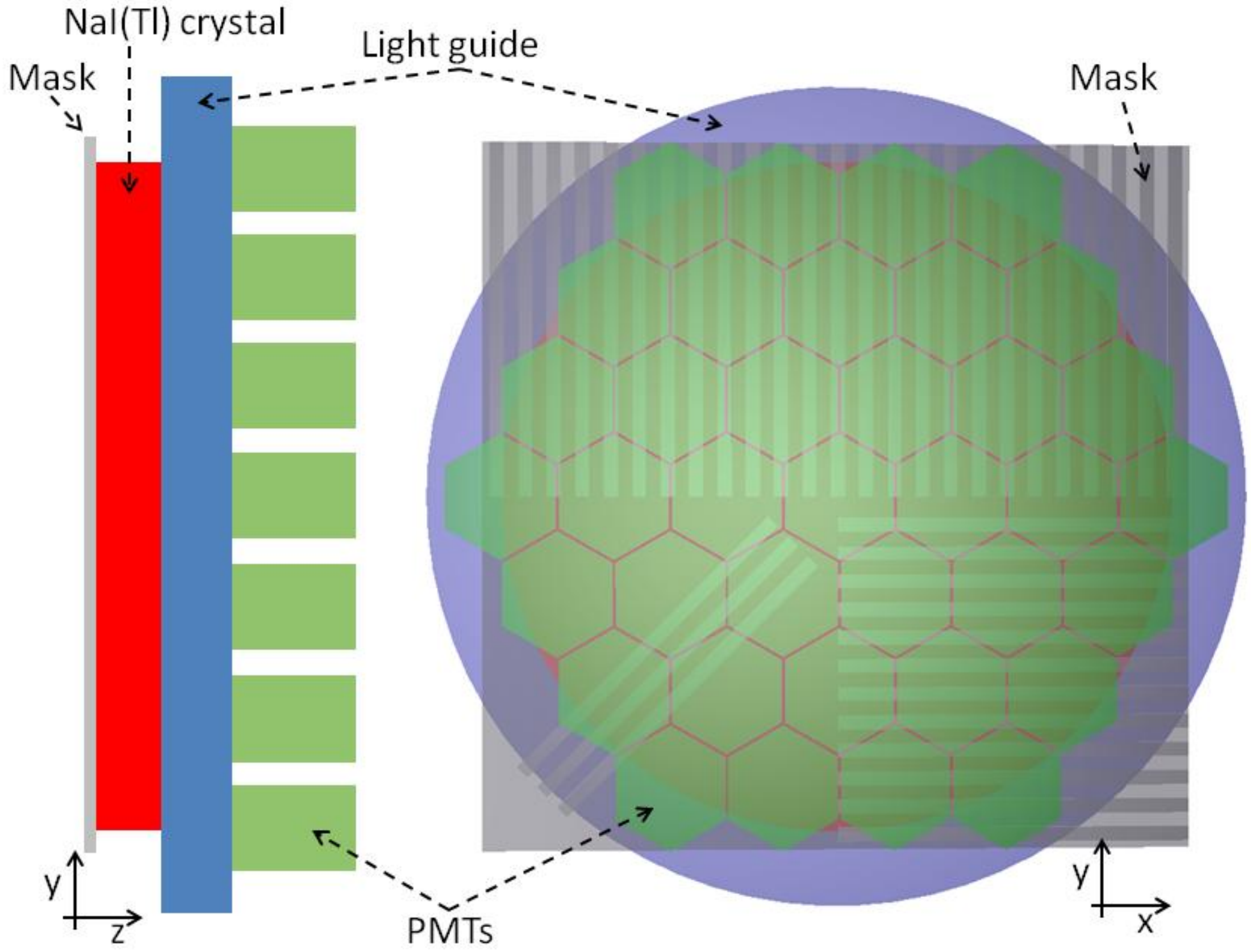

**Figure 1**: Sketch of the medical gamma camera used in simulations. Left: side view, the components are not to scale in z-direction. Right: semitransparent top view. 37 hexagonal PMTs (green) are coupled to the scintillation crystal (red) through a larger diameter light guide. The crystal is irradiated with a parallel beam of gamma rays through a mask of regularly-spaced lead bars positioned behind the crystal.

Typically, $3{\times}10^5$ events resulting in energy deposition in the crystal were simulated. Position reconstruction for every event was performed in two steps. First an approximate position of the event was estimated using the centroid reconstruction. Next, the contracting-grids algorithm, described in section 5, was used for a more accurate evaluation.

In this study we assume axial symmetry of the PMT response, so that each LRF depends only on the distance to the corresponding PMT center. Cubic B-splines [17] with 10 nodes were used for the LRF parameterization. The validity of the assumption on the axial symmetry and the adequateness of the parameterization scheme is demonstrated in figure 2, where the density map of the reconstructed positions of the events (left) and the distribution of the deviation of the reconstructed X-coordinate from its true value (right) are shown for the simulated events reconstructed using the actual LRFs. The figure shows that the image of the mask is practically distortion-free and the 3.6 mm FWHM of the deviation between the reconstructed and true coordinates matches the expected spatial resolution of ~4 mm for this type of gamma cameras.

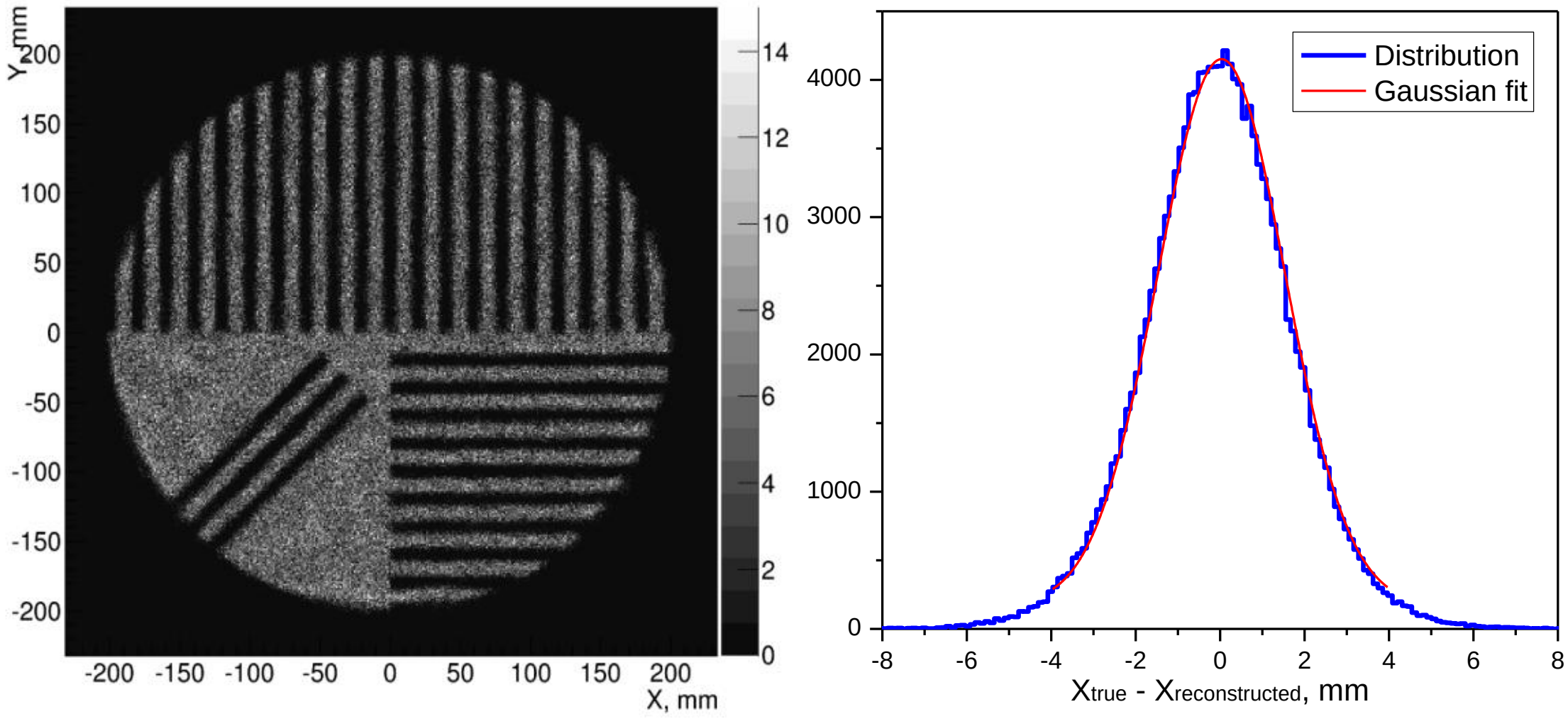

**Figure 2:** Left: Density map (grey scale coded) of the reconstructed positions for a simulation of the gamma camera with the regular mask. The LRFs were calculated using the known positions of the simulated events. Right: Distribution of the difference between the true and the reconstructed X coordinate of the events and a Gaussian fit resulting in 3.6 mm FWHM.

### 3.3 Iterative LRF reconstruction using simulation data

To evaluate the feasibility of the iterative LRF reconstruction technique, the procedure described in section 2 was applied to a dataset of $3\times10^5$ events using several initial guesses. The profiles of the initial LRFs were intentionally chosen to be significantly different from the profiles of the actual LRFs as can be seen at the example shown in figure 3 (top-left). The initial sets of the PMT gains were assigned randomly in the range from 0.25 to 1. Predictably, the image of the mask, reconstructed using the initial LRFs was showing a very strong level of distortions (figure 4, top-left).

Already the first iteration tended to yield LRF profiles much closer to the one of the actual LRF (figure 3, top-right). The following iterations resulted in a gradual improvement of the reconstructed image quality finally reaching the steady state after typically ~40 iterations, with LRFs converged to a profile essentially matching the one of the actual LRFs (see figures 3 and 4). Note that the small difference between the reconstructed and actual LRFs observed close to the zero distance has a very little impact on the reconstruction quality since in the regions close to the PMT centers the reconstructed position is mostly defined by the LRFs of the neighboring PMTs.

The techniques described in section 2 were actively used to improve convergence of the iterative process. The blurring procedure was always applied to the positions reconstructed using the initial guess of the LRFs. Typically, first 10 to 12 iterations were performed using a common LRF for all PMTs until the convergence was reached and then the following iterations were done with individual LRF assigned to each PMT.

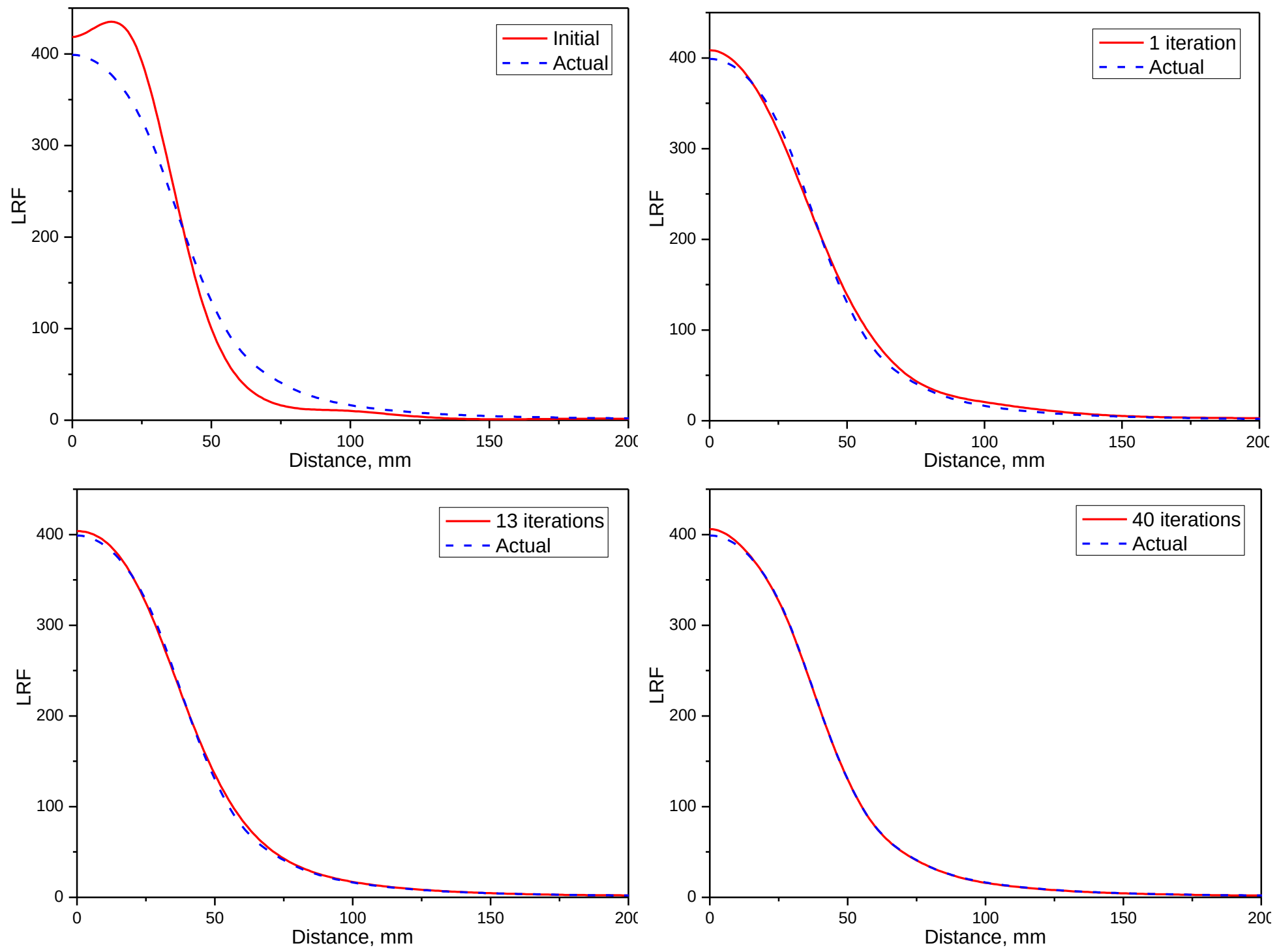

**Figure 3:** LRF of the central PMT obtained during the iterative reconstruction (initial conditions as well as first, 13th and 40th iteration: see legends). Reconstructed LRFs are scaled to have the same value as the actual LRF at the distance of 40 mm.

A similar study was conducted using event data obtained from a simulation in which the NaI(Tl) scintillation yield was artificially reduced by a factor of two to better reproduce our experimental conditions: due to the limited integration time of the data acquisition system, only approximately half of the scintillation photons was collected (see section 4). The results show that the iterative LRF reconstruction can still be successfully applied at these conditions. The reconstructed LRF profiles are nearly identical to the profiles of the actual LRFs and the density maps of the reconstructed positions show only minor distortions (see figure 5). As expected, the apparent spatial resolution deteriorates: the images of the lead bars are clearly broader.

### 3.4 Effective positions of the PMT centers

In order to calculate the expected signal of a PMT for a given light source position, the PMT center has to be defined, which serves as the origin of the LRF function of this photomultiplier. The natural choice is the geometrical center of the PMT window. However, for a practical detector, this position may not coincide with the centroid of the photocathode sensitivity, which, in fact, should be used as the origin of the LRF.

The effective center positions of the PMTs can be determined experimentally by scanning the field of view of the detector with a point gamma source. However, this procedure requires special equipment and is time consuming. We have found that if the differences between the assumed positions (e.g. centers of the PMT windows) and the true center positions are small (on the order of the spatial resolution of the camera), the effective center positions can also be evaluated using a flood source. Two techniques were developed for that purpose.

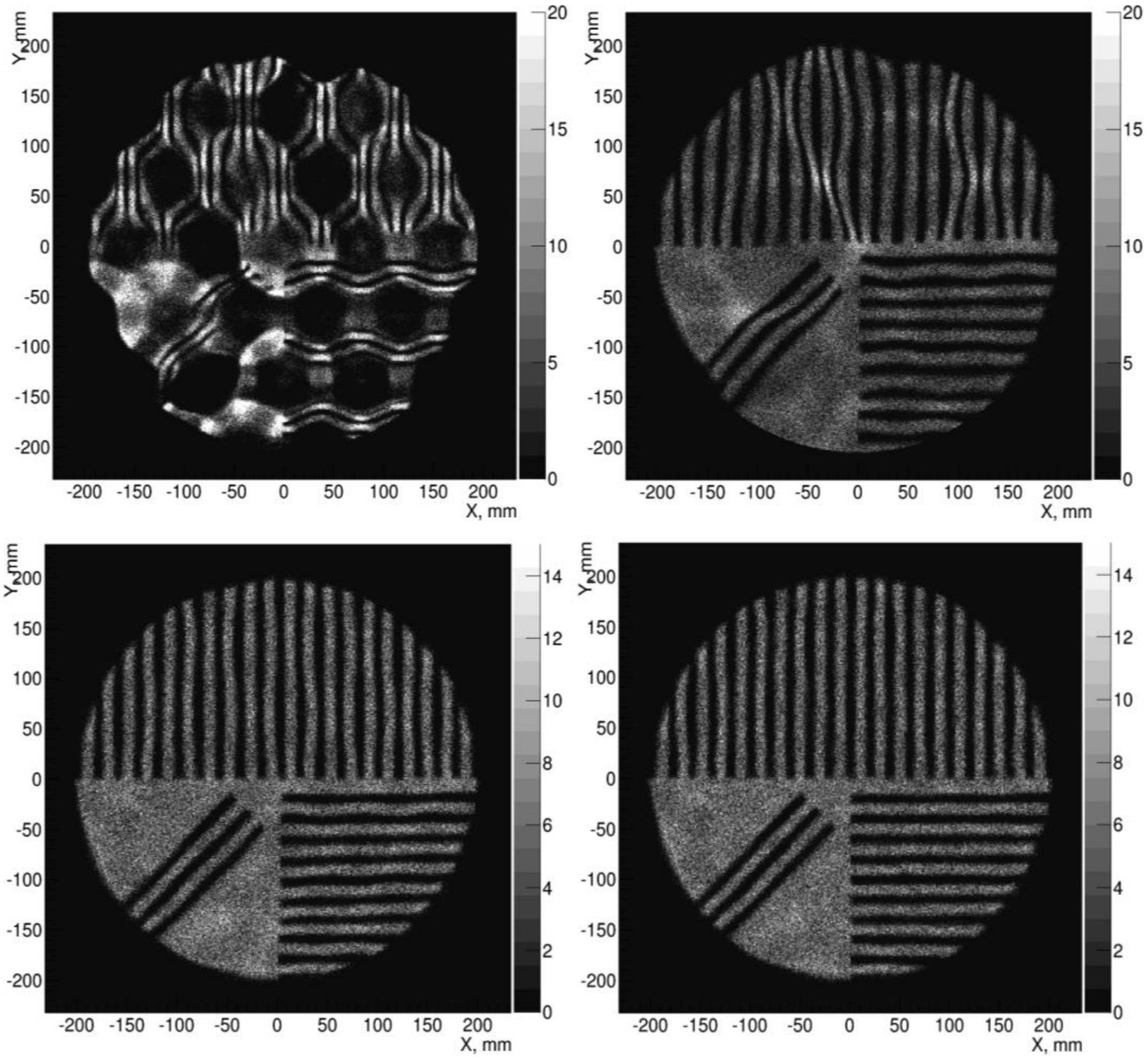

**Figure 4:** Density maps of the reconstructed positions. Top-left: simulated LRFs were used; top-right: one iteration; bottom-left: 12 iterations; bottom-right: 40 iterations.

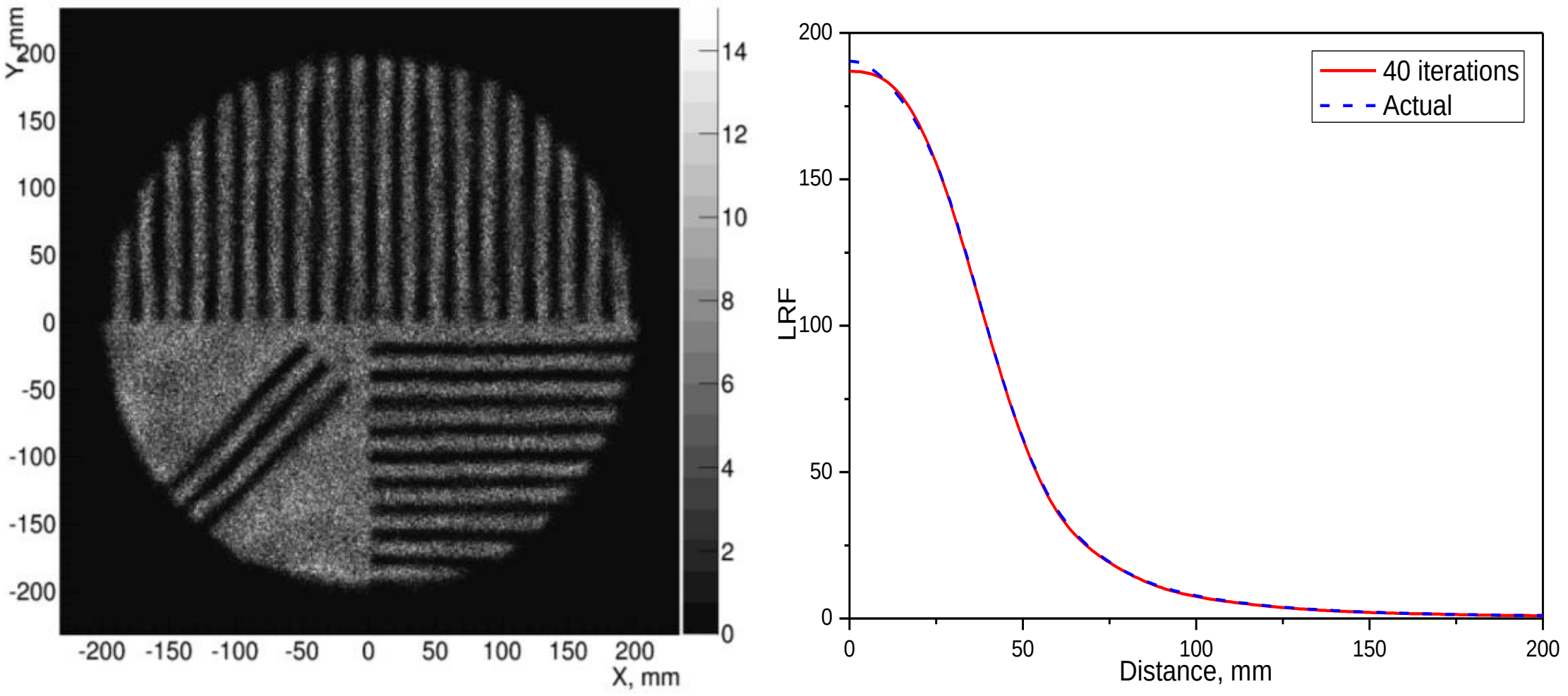

**Figure 5:** Density map of the reconstructed positions (left) and the reconstructed LRF, compared to the actual LRF (right) after 40 iterations for a simulation performed with the number of photons per event reduced by a factor of two.

The first technique (*brute force* approach) consists in the following procedure. A regular grid is defined over the region where the search of the effective center position is to be conducted. For each node of the grid, the same set of events recorded under flood irradiation is reconstructed assuming that the effective center of a selected PMT is at that node position, and the average chi-squared of the reconstruction over the entire dataset is calculated. The estimate of the effective center position of this PMT is given by the node position which results in the smallest average chi-squared.

In the second technique (*Gauss* method), the effective center position of a PMT is found in the following two steps. First a reconstruction of a set of events recorded with flood irradiation is performed without taking the signals of this PMT into account (PMT is "disabled"). Assuming that the effective center positions of the neighboring PMTs are known, the reconstruction of the events situated in front of the disabled PMT is not distorted by the shift in its position. The effective center is then given by the position of the axis of symmetry of the disabled PMT's signal as a function of the reconstructed position. The position of the axis can be found, for example, by fitting the spatial dependence of the PMT signal with a Gaussian profile.

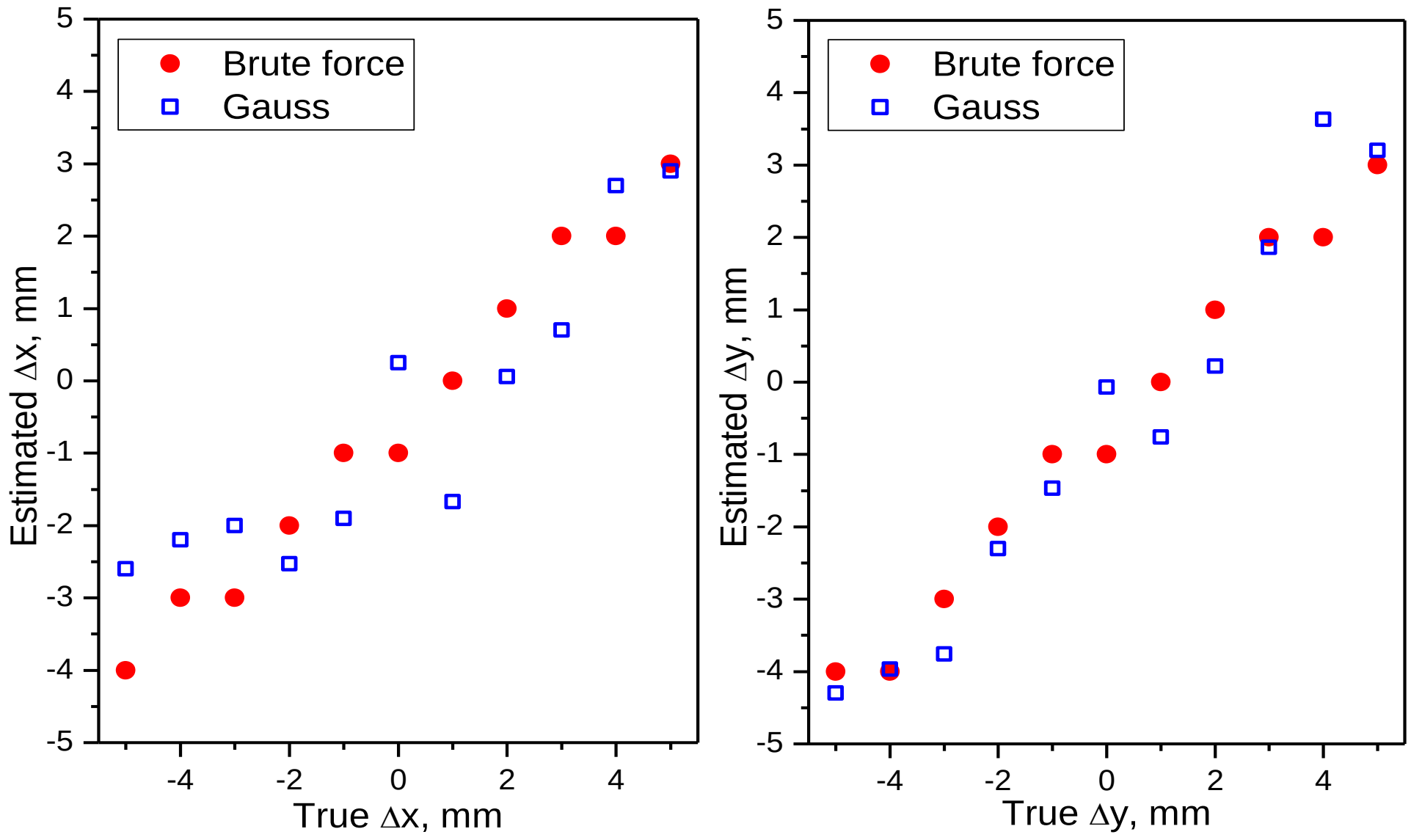

**Figure 6:** Effective center position determination using *brute force* and *Gauss* methods. The estimated offset of the PM center is shown as a function of the true offset.

In order to check the feasibility of these methods, a series of simulations was performed in which the center of one PMT was shifted by a small distance ranging from -5 to +5 mm in X and Y directions (uncorrelated). It was assumed that all PMTs are ideal, i.e. have uniform sensitivity over the photocathode and that the centroid of the sensitivity coincides with the geometric center of the PMT. The iterative reconstruction of the LRFs was performed for all datasets with the initial LRFs obtained from the simulation data for the configuration with zero shifts. The average chi squared for the *brute force* method resulted in a smooth function of coordinates for all datasets. The estimated shift in the PMT center as a function of the true shift is shown in figure 6. The results demonstrate that the true position of the PMT center can be determined with ≤ 2 mm precision. The *brute force* method gives somewhat more accurate results but the procedure is very demanding in terms of calculation power and, therefore, is not recommended if GPU-based event reconstruction (see section 5) is unavailable.

## 4. Validation with experimental data

### 4.1 Experimental setup

An experimental validation of the iterative reconstruction technique described above has been performed using a decommissioned medical gamma camera GE MaxiCamera 400T. This camera has a simple classical design. A 470 mm diameter, 12.5 mm thick NaI(Tl) scintillation crystal is sealed in an aluminium container with a glass window. The crystal is viewed by 37 hexagonal photomultipliers through the window and a plastic light guide of approximately the same thickness as the crystal. The light guide and the PMT array cover a round area with 500 mm diameter. The detector is contained in a lead shielding and is equipped with a lead collimator. A standard high resolution low energy collimator (2.5 mm diameter round holes, 0.5 mm septa) was used in the present work. The collimator limits the detector field of view by a circle with the diameter of 400 mm.

Originally, the signals of each PMT were fed to individual amplifiers and then injected to an analogue resistive circuit, where the (x,y) positions of scintillation events were reconstructed using the center of gravity method. The circuit was mounted inside the lead case so that only X+, X-, Y+, Y- and E (sum of all PMT signals) were passed to the exterior. The simplicity of the classical design allowed us to easily access the signals of the individual PMTs, necessary for implementation of the statistical event reconstruction. The PMT signals, read on the anode, were connected to an adapter board that allowed to connect the camera to the external signal processing and data acquisition system.

The acquisition system used in our experiments was based on the MAROC3 ASIC [18]. The MAROC3 front-end section features 64 low noise inputs with individually adjustable gains and two (slow and fast) signal processing chains. The slow signal was used for amplitude measurements being digitized by the on-chip 12 bit Wilkinson ADC. The integration time of the slow shaper was set to 150 ns, which is the maximum time allowed by the MAROC3 chip. Unfortunately, this integration time resulted in suboptimal (about 50%) charge collection, as the decay time for NaI(Tl) scintillator is ~250 ns. The raw ADC data were transferred to a PC, where a pre-processing procedure was applied to subtract the pedestals and to filter out the saturated signals due to the background radioactivity.

The trigger was also obtained from the MAROC3 chip. The outputs of the preamplifiers were all summed together and the resulting sum signal was fed into one of the MAROC3 channels and was used to trigger the system. The discrimination level was set low enough to guarantee triggering from the 122 keV events even on the periphery of the crystal. The channel gains were adjusted to make triggering efficiency as uniform as possible over the field of view of the detector. The fact that MAROC3 permits digital control over its front-end configuration allowed us to realize automatic adjustment of the gains by software. The acquisition rate was limited to ~3 kHz which was considered to be sufficient for research purposes.

### 4.2 Iterative LRF reconstruction using experimental data

Experimental data were recorded using a $^{57}$Co flood gamma-ray source (122 keV and 136 keV with 89% and 11% probability, respectively) placed parallel to the collimator surface at a distance of 300 mm. A mask made of 3 mm thick, 200 mm long and 10 mm wide lead bars spaced by 10 mm was installed in contact with the collimator. The mask pattern was very similar to the one used in the simulations (see section 3.2). The source provided a uniform irradiation of the entire detector viewing area allowing to use the reconstructed image of the mask for evaluation of the distortions introduced by the reconstruction procedure.

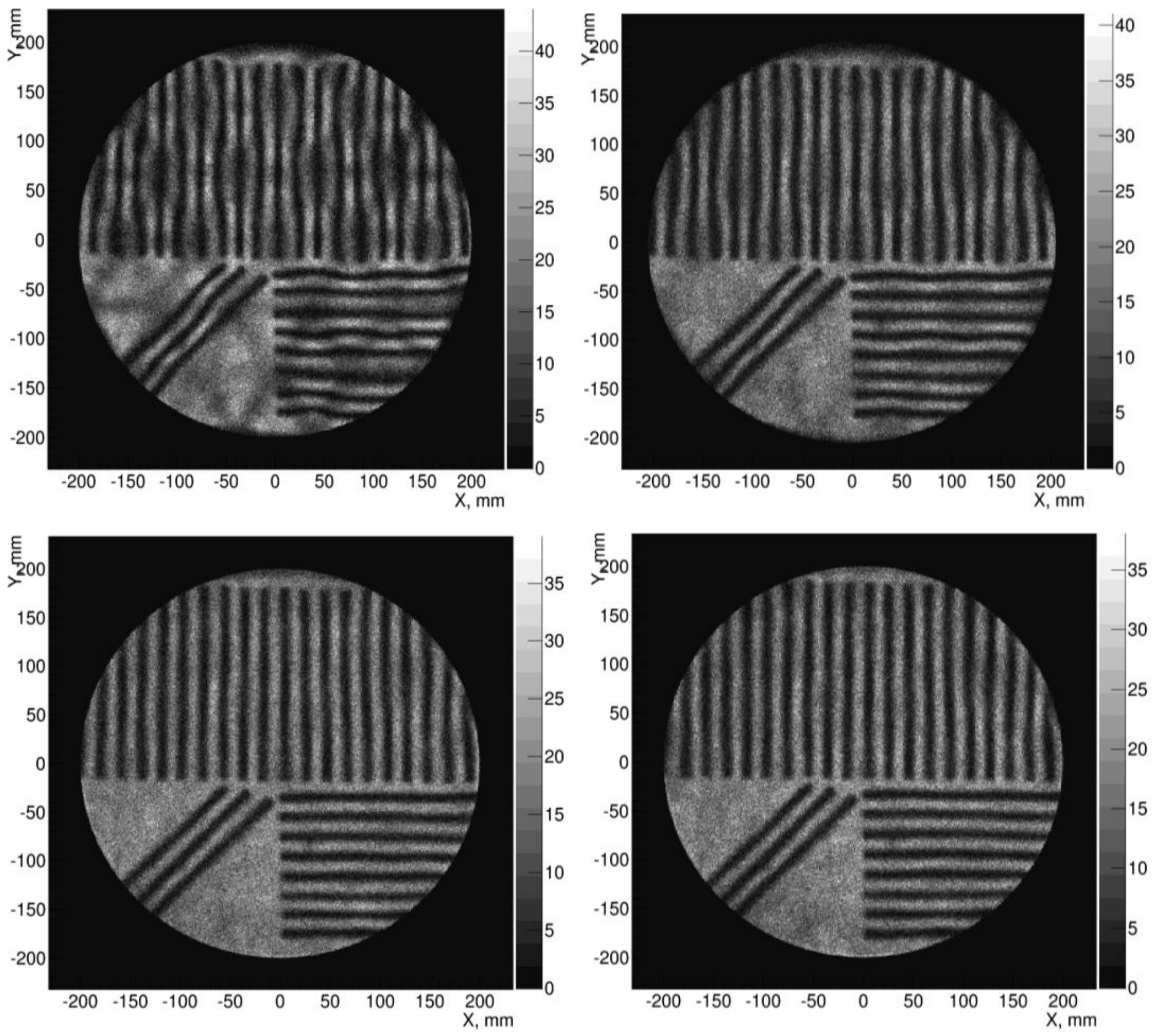

**Figure 7:** Density maps of the reconstructed positions obtained during the iterative LRF reconstruction. Top-left: LRF from simulations (initial estimation); Top-right: second iteration, common LRFs; Bottom-left: after 10 iterations, individual LRFs; Bottom-right: after 40 iterations.

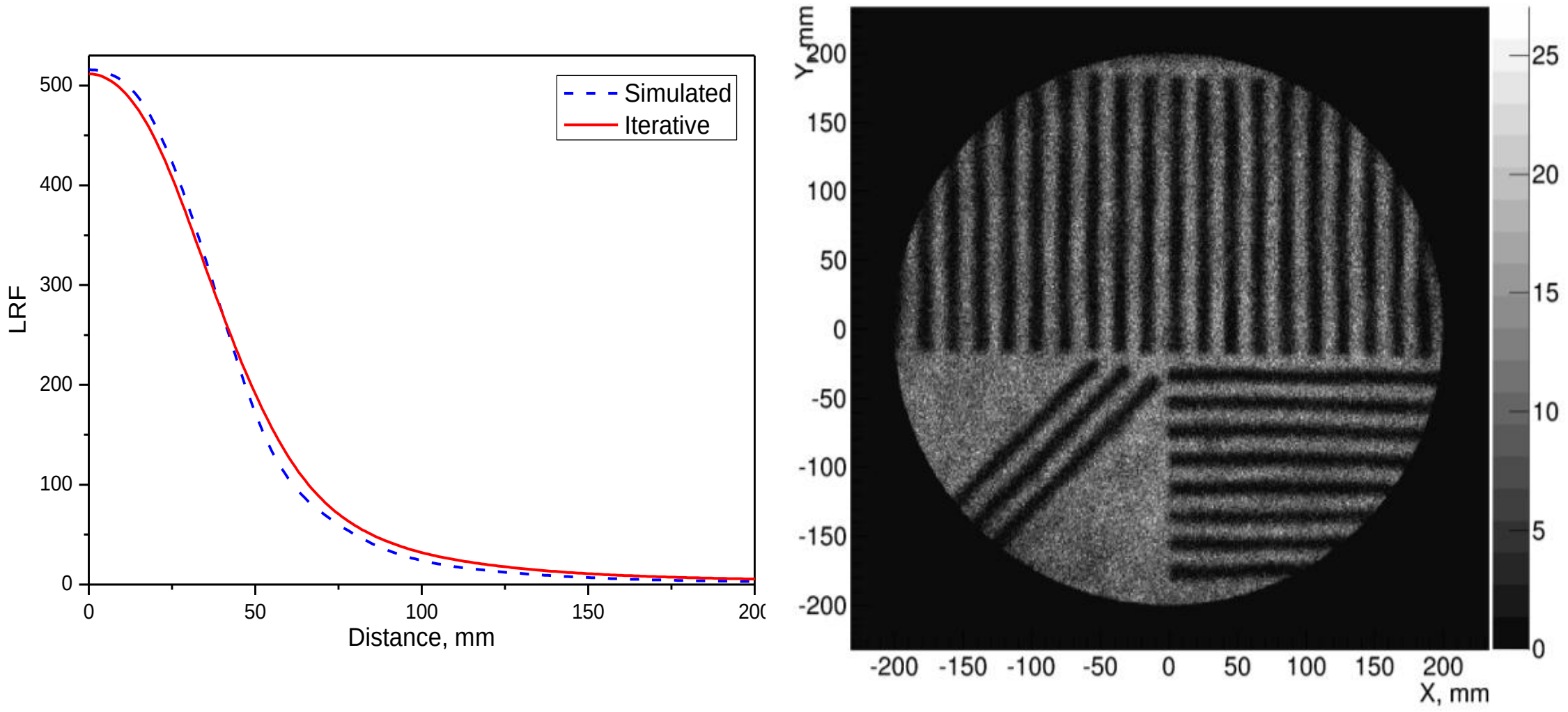

**Figure 8:** Left: Initial LRF (dashed line), calculated using simulated data and the final LRF (solid line) provided by the iterative LRF reconstruction technique. The example is given for PMT #20 (see figure 11). The final LRF is scaled so that both LRFs have the same value at the distance of 40 mm. Right: Density map of the reconstructed positions after correction of the PMT #13 center position and two additional iterations.

The initial estimation of the LRFs was taken from the simulations, assuming equal gains for all PMTs. As described in section 3, the LRFs were parameterized as radial functions using cubic B-splines with 10 nodes. Reconstruction of the experimental data with these initial LRFs resulted in the density map shown in figure 7 (top-left). The distortions are relatively small which indicates that the simulation model is quite adequate for this camera.

The iterative LRF reconstruction was performed following the procedure described in section 3.3, which also provided the fastest convergence for the experimental data. The first nine iterations were done using a common LRF for all PMTs. Starting from the 10th iteration, individual LRFs were used. The steady state was typically reached after the following 30 iterations. An example of the resulting density map is shown in figure 7 (bottom-right), and the difference between the initial LRF and the final LRF at the steady state for one of the PMTs is given in figure 8 (left).

The final density map shows a low level of distortion everywhere except for a region around the point with (100,100) coordinates (see figure 7, bottom right). The center of the distorted region is situated in front of the PMT #13 (see PMT numbering scheme in figure 11), thus suggesting that there may be a significant difference between the effective sensitivity center of that PMT and its position assumed in the reconstruction. An evaluation of the effective center position of this PMT was performed as described in section 3.4. Both methods indicated that the center is shifted by -4 mm along the Y axis. After the center coordinates were adjusted and two additional iterations were performed, the reconstruction pattern showed significantly less distortions in the vicinity of this PMT (figure 8, right).

Analysis of the X-projection of the upper half of the density map shows that the positions of the lead bars are properly reconstructed: note the 20 mm step in figure 9 (left). The energy spectrum (figure 9, right) exhibits a peak from $^{57}$Co source with a FWHM of ~14%. The width is somewhat broader than the value listed in the specifications of the gamma camera (11% for $^{57}$Co source), which is expectable since the acquisition system used in these measurements collects only about one half of the total number of the scintillation photons emitted by the NaI(Tl) crystal. As shown in the simulations (section 3.3), we expect an improvement in the reconstruction quality and better spatial resolution for a system with a longer integration time.

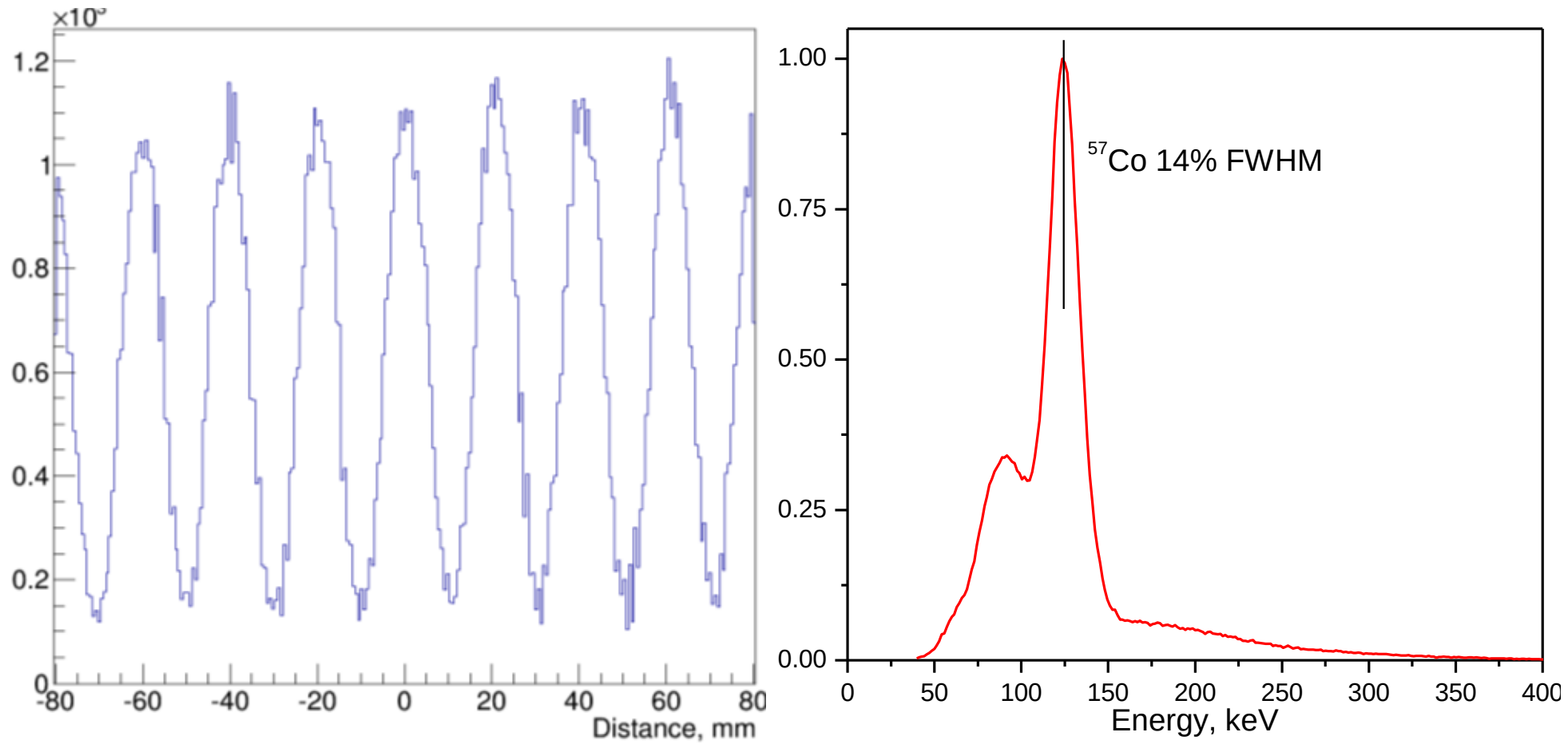

**Figure 9:** Left: X-projection of the reconstructed event density (Y range from -15 to 200 mm). The peak positions reproduce well the 20 mm pitch of the mask. Right: Reconstructed energy spectrum showing the 122 keV $^{57}$Co peak and a continuum due to the background radioactivity. The 122 keV peak has a FWHM of ~14%.

### 4.3 Gain monitoring

Iterative LRF reconstruction technique can also be used to monitor fluctuations of PMT gains and correct them if necessary. In order to demonstrate this, we have used the fact that MAROC3 readout system has adjustable gain for each individual channel. Below we refer to these gains as *hardware* gains.

The capability to accurately evaluate gain of a single PMT was shown by the following experiment. The hardware gain of one MAROC3 channel was varied in the range from 3 to 60 (see figure 10) and a set of $3\times10^5$ events was recorded for each gain value. The events were processed using the iterative LRF reconstruction technique with grouping of all PMTs, providing a common LRF and relative gains of all the channels for each event set. The result is shown in figure 10 where one can see a linear correlation between the manually set hardware gain and the reconstructed gain value. For the channels with fixed hardware gains, the reconstructed gain values were practically the same for all data sets.

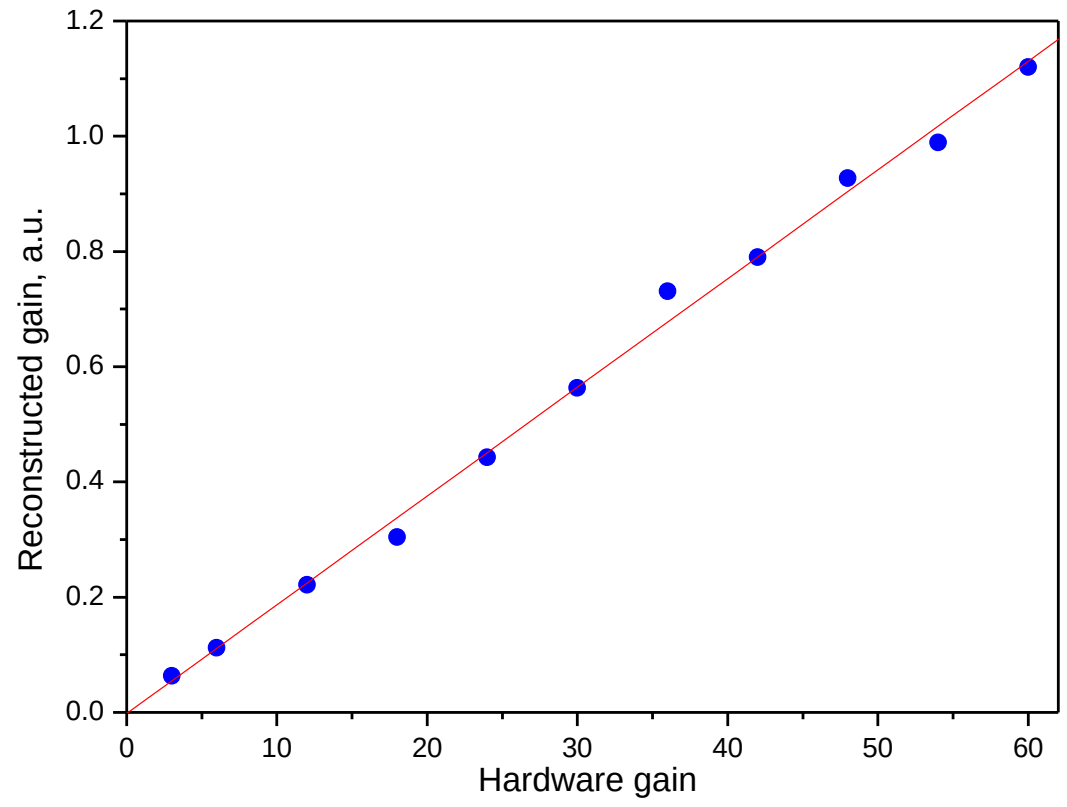

**Figure 10:** Reconstructed gain as a function of the manually set hardware gain in one of the MAROC3 channels.

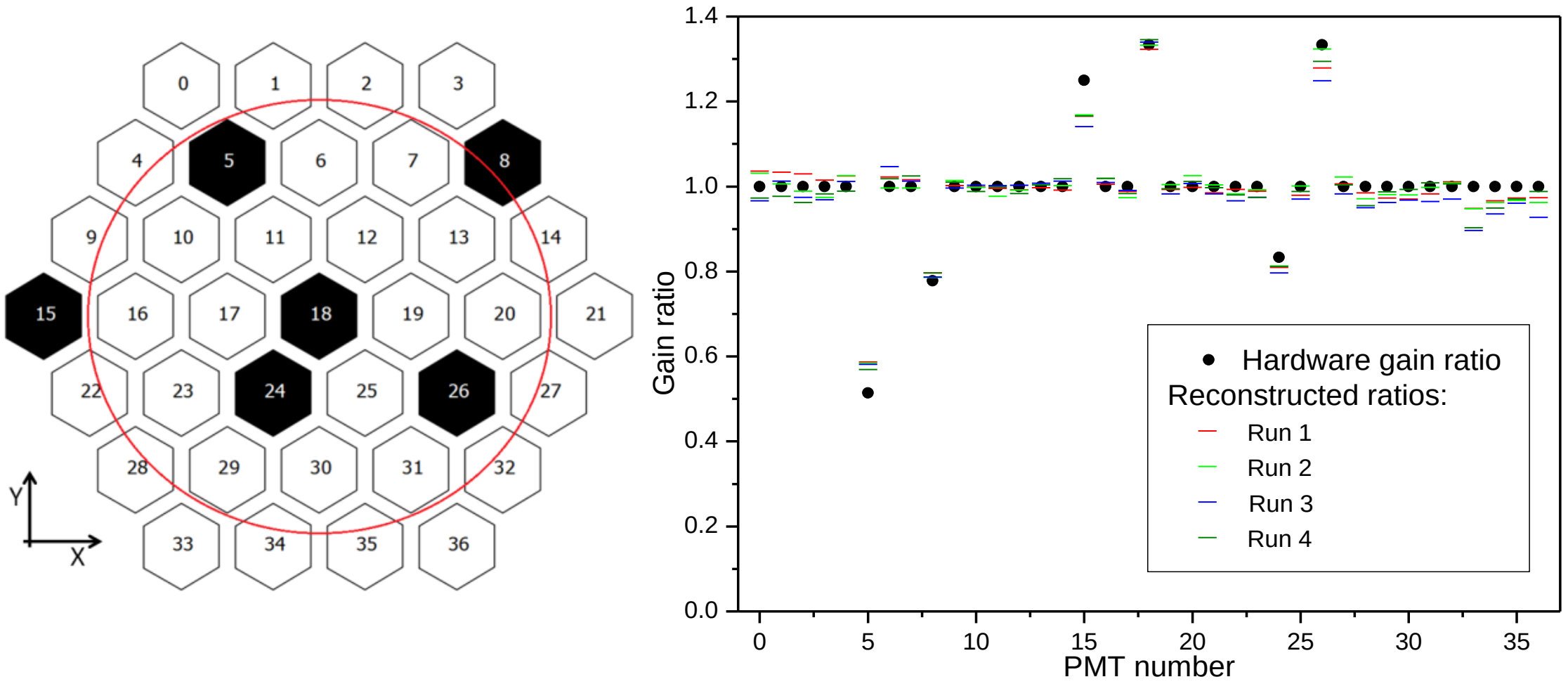

**Figure 11.** Left: The array of photomultipliers and six PMTs (black filled) with manually modified hardware gains. The red circle shows the detector's sensitive region limited by the collimator. Right: The ratio of the known hardware gains (black dots) and the gain ratios calculated using the iterative LRF reconstruction for four independent runs (horizontal bars).

The capability of our method to evaluate simultaneous gain changes in several PMTs was also investigated. A set of $1\times10^5$ events was acquired, then the hardware gains of six PMTs (see figure 11, left) were manually modified and another set of $1\times10^5$ events was recorded. The ratios of the hardware gains (the initial values to the modified ones) are shown in figure 11 (right) with black dots. Iterative LRF reconstruction was then performed for both sets of events providing reconstructed gain values. The ratios of the reconstructed gains for the initial and the modified configurations were calculated and compared with the known hardware gain ratios.

This procedure was repeated three more times. The obtained gain ratios are shown in figure 11 (right) with horizontal bars. The four independent runs demonstrate a good agreement, and the differences between the hardware gain ratios and the corresponding reconstructed ratios are quite small. Note that the largest deviations are obtained for the PMTs which are situated entirely outside the detector field of view (see figure 11, left).

## 5. Real-time implementation

One of the practical reasons why statistical reconstruction algorithms are used much less frequently than the center of gravity reconstruction is the high calculation cost of such algorithms. Until recently, statistical reconstruction could not handle the event rates at which medical gamma cameras are typically operated (~100 kHz). Here we show that with the LRF parameterization scheme proposed in this work and using parallel calculations on a GPU it is possible to reach the required real-time reconstruction rate.

In this study, position reconstruction of all events was performed with a modified contracting-grids algorithm [19]. The algorithm works in the following way. For each event a regular grid of positions is defined, which is centered at the (x, y) coordinates given by the centroid reconstruction of this event. For each node, the algorithm evaluates the difference between the measured (or simulated) PMT signals and the corresponding expected signal values given by the LRFs assuming that the light was emitted from this node. The evaluation is performed using the least squares approach, and the grid node resulting in the smallest chi-squared is selected and used as the center of a finer grid covering the vicinity of this position. The procedure is repeated until a sufficiently small grid step (and, hence, the reconstruction precision) is reached. The event energy at each node is estimated using the analytical expression given by equation 5 in [13].

This algorithm was implemented on a GPU using CUDA platform. A CUDA thread block was allocated for reconstruction of each event. Every node of the grid was processed by an individual thread within the thread block by performing the calculation of the chi-squared, assuming that the event position is at that node. The node showing the smallest chi-squared was selected and a new search with a finer grid centered on this node was performed using the same thread block. The number of grids used in the search was predefined by the user.

Since the B-spline parameterization of LRFs results in very compact data (typically less than 20 float values per LRF), this information can be stored in the CUDA constant memory allowing very fast access. Using a consumer grade PC equipped with a 3.4 GHz Inter Core i7 processor and NVIDIA GTX 770 GPU board, the average reconstruction time per event was approximately 2 μs. Half of this time was the actual calculation on the GPU (CUDA kernel execution time), and the other half was spent on data preparation/transfer and event filtering. The optimal performance was reached with the contracting-grids algorithm using three iterations on 8x8 grids. The overall reconstruction rate of $5\times10^5$ events per second can be improved for a specific practical application sacrificing the generality of the algorithm developed in the frame of the ANTS2 package.

The results of the reconstruction performed with the contracting-grids algorithm were compared with those obtained with another least squares technique in which chi-squared minimization over spatial coordinates and event energy was performed using the Migrad algorithm from Minuit2 library (available in ROOT). Both algorithms provided nearly identical reconstruction results.

Very fast reconstruction also benefits the iterative LRF reconstruction procedure: with $3\times10^5$ events using 10 nodes B-spline LRF parameterization, it takes only a few seconds to perform a complete iteration cycle (position reconstruction followed by the calculation of LRFs) on the PC described above. Even with 60 iterations (usually less is required), the entire procedure is completed within less than five minutes.

## 6. Conclusions

In this study we have demonstrated the feasibility of the iterative LRF reconstruction technique for medical gamma cameras. The technique provides accurate light response functions of the photomultipliers using only flood field irradiation data. The feasibility has been shown both with simulated and experimental data recorded with a retrofitted commercial gamma camera.

It was found that for the gamma camera used in the study the assumption on axial symmetry of the LRFs is adequate and that the radial dependence of LRFs can be successfully parameterized as a weighted sum of cubic B splines. We have also shown that the iterative reconstruction technique, combined with a GPU-based event reconstruction can be used for real-time processing of the events as well as for continuous on-the-fly calibration of the PMT gains.

A successful application of the iterative technique to the data recorded with a flood source and a lead mask demonstrates that the method does not have strict requirements on the uniformity of the detector irradiation. This fact points to a possibility of camera calibration with the data recorded from a patient.

## Acknowledgments

This work was carried out with financial support from Fundação para Ciência e Tecnologia (FCT) through the project grant PTDC/BBB-BMD/2395/2012 (co-financed with FEDER) and from Quadro de Referência Estratégica Nacional (QREN) in the framework of the project Rad4Life.